\documentclass[journal]{IEEEtran}
\usepackage{amsmath}
\usepackage{multirow}
\usepackage{graphicx}
\usepackage{cite}
\usepackage{subfigure}
\usepackage{amssymb}
\usepackage[urlcolor=blue]{hyperref}
\makeatletter
\newcommand*\bigcdot{\mathpalette\bigcdot@{.5}}
\newcommand*\bigcdot@[2]{\mathbin{\vcenter{\hbox{\scalebox{#2}{$\m@th#1\bullet$}}}}}
\makeatother
\ifCLASSINFOpdf
\else
\fi

\begin{document}
\title{MSStyleTTS: Multi-Scale Style Modeling with Hierarchical Context Information for Expressive Speech Synthesis}

\author{Shun Lei,~\IEEEmembership{Student Member,~IEEE,}
        Yixuan Zhou,~\IEEEmembership{Student Member,~IEEE,}
        Liyang Chen,~\IEEEmembership{Student Member,~IEEE,}
        Zhiyong Wu,~\IEEEmembership{Member,~IEEE,}
        Xixin Wu,~\IEEEmembership{Member,~IEEE,}
        Shiyin Kang,~\IEEEmembership{Member,~IEEE,}
        Helen Meng,~\IEEEmembership{Fellow,~IEEE}

\thanks{This work was supported by National Natural Science Foundation of China (62076144), Shenzhen Science and Technology Program (WDZC20220816140515001), Shenzhen Key Laboratory of next generation interactive media innovative technology (ZDSYS20210623092001004) and AMiner.Shenzhen SciBrain fund. \textit{(Shun Lei and Yixuan Zhou contributed equally to this work. Corresponding author: Zhiyong Wu.)}}
\thanks{Shun Lei, Yixuan Zhou, Liyang Chen and Zhiyong Wu are with Tsinghua-CUHK Joint Research Center for Media Sciences, Technologies and Systems, Shenzhen International Graduate School, Tsinghua University, Shenzhen, China (e-mail: leis21@mails.tsinghua.edu.cn; zhouyx20@mails.tsinghua.edu.cn; cly21@mails.tsinghua.edu.cn; zywu@ sz.tsinghua.edu.cn). This work was conducted when Shun Lei was an intern at Xverse Inc.}
\thanks{Xixin Wu and Helen Meng are with Department of Systems Engineering and Engineering Management, The Chinese University of Hong Kong, Hong Kong SAR, China (e-mail: wuxx@se.cuhk.edu.hk; hmmeng@se.cuhk.edu.hk).}
\thanks{Shiyin Kang is with Skywork AI Pte. Ltd., Beijing, China (e-mail: shiyin.kang@ushow.media).}
}

\markboth{Journal of \LaTeX\ Class Files,~Vol.~14, No.~8, August~2021}%
{Shell \MakeLowercase{\textit{Shun Lei et al.}}: MSStyleTTS: Multi-scale Style Modeling with Hierarchical Context Information for Expressive Speech Synthesis}

\maketitle

\begin{abstract}
Expressive speech synthesis is crucial for many human-computer interaction scenarios, such as audiobooks, podcasts, and voice assistants.
Previous works focus on predicting the style embeddings at one single scale from the information within the current sentence.
Whereas, context information in neighboring sentences and multi-scale nature of style in human speech are neglected, making it challenging to convert multi-sentence text into natural and expressive speech.
In this paper, we propose MSStyleTTS, a style modeling method for expressive speech synthesis, to capture and predict styles at different levels from a wider range of context rather than a sentence.
Two sub-modules, including multi-scale style extractor and multi-scale style predictor, are trained together with a FastSpeech 2 based acoustic model.
The predictor is designed to explore the hierarchical context information by considering structural relationships in context and predict style embeddings at global-level, sentence-level and subword-level.
The extractor extracts multi-scale style embedding from the ground-truth speech and explicitly guides the style prediction.
Evaluations on both in-domain and out-of-domain audiobook datasets demonstrate that the proposed method significantly outperforms the three baselines.
In addition, we conduct the analysis of the context information and multi-scale style representations that have never been discussed before.

\end{abstract}


\IEEEpeerreviewmaketitle

\ifCLASSOPTIONcaptionsoff
  \newpage
\fi

\section{Introduction}
\label{sec:introduction}
\IEEEPARstart{T}{ext-to-speech} (TTS), which aims to generate natural and intelligible speech from text, has widespread applications in human-computer interaction.
The traditional TTS methods include concatenative synthesis \cite{olive1977rule, hunt1996unit, campbell1997prosody, wang2000corpus, black1997automatically} and statistical parametric speech synthesis \cite{yoshimura1999simultaneous, tokuda2000speech, zen2009statistical, tokuda2013speech, ze2013statistical}.
With the rapid development of deep learning, neural network based TTS models \cite{sotelo2017char2wav, tacotron, tacotron2, deepvoice3,li2019neural, Yu2020DurIANDI, fastspeech2} are devised to generate spectrograms directly from phoneme sequences, which brings great progress in synthesizing high-quality speech with a neutral style.
However, there is still a clear gap between synthesized speeches and human recordings in terms of expressiveness, which hinders the advancement of speech synthesis technology.
One important factor is that there are complex and multi-scale stylistic variations in human recordings, affected by multiple factors (including contextual information, speaker’s intention, etc.), causing difficulties for TTS models to learn directly from phoneme sequences.

To model the style,
one general approach is style transfer TTS \cite{reference, gst, VAEGST, an2019learning, wu2019end,hsu2018hierarchical, multiscale, yi2022prosodyspeech}.
It is designed to generate speech with the same style as the given reference audio.
In \cite{reference}, a reference encoder is trained to obtain the style embedding at global level from the given speech.
The global style token (GST) and their derivatives further utilize several learnable style tokens to represent the global style, and perform well on the style transfer at sentence-level \cite{gst, VAEGST, wu2019end, an2019learning}.
Other recent proposed methods seek to control the local prosodic characteristics by considering fine-grained style representations \cite{lee2019robust, yi2022prosodyspeech, klimkov2019fine}.

Compared with style transfer TTS that require auxiliary inputs in inference, the approach that directly predicts style from text is more practical and flexible.
The text-predicted global style token (TP-GST) model \cite{TPGST} firstly introduces the idea of predicting style embedding or style token weights from input phoneme sequence.
Speeches with more pitch and energy variations than baseline Tacotron \cite{tacotron} can be generated in this way.
Considering that style and semantic information of utterance are closely related, the text embeddings derived from pre-trained language models (PLMs), e.g., Bidirectional Encoder Representations from Transformer (BERT) \cite{bert}, have been incorporated to TTS models to improve the expressiveness of the synthesized speech \cite{berttacotron, bertemb, fang2019towards}.
Fine-grained style representations to model local prosodic variations in speech, such as word level \cite{wsv, ren2022prosospeech, liu22m_interspeech} and phoneme level \cite{tan2020fine, du2021rich}, are also considered in some works.

However, the above text-predicted methods have two major limitations.
First, these approaches only take into account the information of the current sentence to be synthesized.
For the same sentence, these models lack the ability to capture the diverse speech variations (e.g., intonation, stress and emotion) that may be affected by its neighboring sentences.
It leads to the lack of coherence between adjacent sentences and accurate expression of synthesized speech.
This is against the human perception that the speech style of the current utterance should be primarily influenced by the context \cite{survey, longformevaluationg, wu2022self}.
In this regard, it has shown that considering a wider range of context is helpful for expressive speech synthesis \cite{nakata11audiobook, crossutterance, li2022cross,xue2022paratts}.
Second, these methods only focus on modeling the mono-scale style representations of human recordings. 
However, the style expressiveness of human recordings varies from coarse to fine granularity \cite{selkirk1986derived, liberman1977stress, tseng2005fluent}.
It shall be regarded as a composition of multi-scale acoustic factors, among which global scale is typically observed as timbre and emotion and is tended to be consistent throughout the entire utterance.
The other is the local scale, which is closer to the prosody variation, including the stress, pause, pitch, and other acoustic features of speech.
These different scales of style work together to produce rich expressiveness in speech.
Towards this end, some recent studies on similar tasks, like emotional speech synthesis \cite{lei2021fine, msemotts} and style transfer \cite{multiscale}, attempt to model stylistic variations at multiple scales.
But these approaches require auxiliary inputs, such as emotion labels and reference speech, and only consider sentence in isolation.
To the best of our knowledge, there is currently no work investigating multi-scale speaking style modeling by considering only text information.

In this paper, we propose MSStyleTTS, a novel multi-scale style modeling method that models styles at global-level, sentence-level, and subword-level\footnote{For Chinese, one Chinese character corresponds to one subword, and one subword is composed of at least one phoneme.} from hierarchical context information for expressive speech synthesis. 
It contains a multi-scale style extractor, a multi-scale style predictor, and an acoustic model based on FastSpeech 2 \cite{fastspeech2}.
The acoustic model synthesized speech of the current sentence with the multi-scale style embedding provided by the extractor or the predictor.
The extractor is used to extract style embeddings at the above three levels from the corresponding ranges of mel-spectrograms, respectively.
It is jointly trained with the acoustic model to learn the multi-scale style representations in an unsupervised way and then explicitly guide the training of the predictor.
Besides, we represent the style variations at different levels as residuals to reduce the overlap of style information between different levels, which make the subsequent prediction task easier.
The predictor generates the style embeddings at these three levels by utilizing a wider range of contextual information.
By utilizing the hierarchical context encoder, the predictor considers both structural relationships in context and the contextual information given by the pre-trained BERT \cite{bert}.
Evaluations on both in-domain and out-of-domain audiobook datasets demonstrate that the naturalness and expressiveness of speeches generated by MSStyleTTS are significantly better than those of baseline FastSpeech 2 \cite{fastspeech2}, WSV* \cite{wsv}, and HCE \cite{proposed} models.
The speech samples generated by our proposed model are available online
\footnote{Speech samples: \href{https://thuhcsi.github.io/TASLP-MSStyleTTS}{https://thuhcsi.github.io/TASLP-MSStyleTTS}}.

Our preliminary work has been presented in \cite{lei22c_interspeech}.
In this study,
the preliminary model is extended with an autoregressive style predictor to effectively capture the multi-scale speaking style and maintain coherence between sentences, which shows the effect on improving paragraph-based speech synthesizing.
We conduct extensive subjective evaluations and ablation studies to demonstrate the effectiveness of the techniques employed in our model.
Moreover, we further analyze how the expressiveness of synthesized speech is influenced by different ranges of context information and different levels of style representations utilized in MSStyleTTS, which has never been studied in previous work.
Overall, the main contributions of this paper are summarized as follows:
\begin{itemize}
    \item This paper presents MSStyleTTS, to our knowledge, which is the first attempt to model the multi-scale style from the context. It significantly improves the expressiveness and coherence of generated speech for both single-sentence and multi-sentence test.
    \item To learn the style representations, we propose a multi-scale style extractor that extracts style embeddings at subword-level, sentence-level and global-level from the corresponding ranges of mel-spectrogram and represents the style variations at different levels as residuals.
    \item For style prediction, we propose a hierarchical context encoder and two hierarchical style predictors to explore the structural relationship in context and predict styles at different levels with residual connections. Moreover, an autoregressive style predictor is designed for achieving paragraph-based synthesis.
    \item Evaluations on both in-domain and out-of-domain datasets show that MSStyleTTS achieves the goal of synthesizing speech with rich expressiveness.
    Extensive ablation studies demonstrate the effectiveness of techniques employed in MSStyleTTS, especially for the context information usage and multi-scale style representation.
\end{itemize}

The rest of the paper is organized as follows. 
We introduce related work in Section \ref{sec:relatedwork},
followed by details of our proposed approach in Section \ref{sec:model}.
Experimental results and ablation studies are presented in Section \ref{sec:exp}.
Section \ref{sec:conclution} gives conclusion of our work.
\section{Related Work} \label{sec:relatedwork}
\subsection{Reference Audio based Style Modeling}
The reference audio based style modeling aims to generate speech with the style that is transferred from a given audio.
In \cite{reference}, an unsupervised reference encoder is utilized to extract sentence-level global style embedding from given audio.
Multiple works have been conducted to study more effective global style representations.
Specifically, the global style could be represented by utilizing a fixed number of learnable global style tokens (GST) \cite{gst}, or by incorporating the variational autoencoder \cite{VAE} to achieve better style disentanglement \cite{VAEGST}, or by using several GST layers with residual connections to learn hierarchical embeddings implicitly \cite{an2019learning}.
In addition, \cite{wu2019end} further introduces an explicit relationship between style tokens and emotion categories in order to enhance the interpretability of the learned style tokens.

Instead of modeling the global style of speech in a sentence level, some researches focus on local prosodic characteristics \cite{lee2019robust, yi2022prosodyspeech, klimkov2019fine, tan2020fine}.
\cite{lee2019robust, yi2022prosodyspeech} introduce the idea of utilizing an additional reference attention mechanism to align the extracted style embedding sequence with the phoneme sequence.
Other studies, such as \cite{klimkov2019fine, tan2020fine}, achieve the same goal based on the forced-alignment technology.
Moreover, a recent study considering the multi-scale nature of speech style \cite{multiscale} also draws our attention.
It proposes a multi-scale reference encoder to extract both the global-scale sentence-level and the local-scale phoneme-level style features from the given reference speech.

In this paper, the proposed method also learns the style representations at different levels. 
In addition to the sentence-level and subword-level, we also consider the global-level that presents the overall style of a wider range of speech containing adjacent sentences.
Furthermore, MSStyleTTS can directly predict the multi-scale style representations from the context without reference audio.
Besides, the residual style embeddings is used to represent style variations at each level rather than using the extracted style embeddings directly, which facilitates the subsequent prediction task.

\subsection{Text Predicted Style Modeling}
It is important to note that during inference stage, auxiliary inputs, such as manually-determined reference audio or token weights, are required for all of the aforementioned researches \cite{reference, gst, VAEGST,wu2019end, an2019learning,lee2019robust, yi2022prosodyspeech, klimkov2019fine, tan2020fine,multiscale}.
To prevent these manual interventions at inference stage, the text-predicted global style token (TP-GST) model \cite{TPGST} extends the GST \cite{gst} by modeling style embedding or style token weights from the input phoneme sequences.
Thus, it is possible to synthesize stylistic speech conditioned on the predicted style embedding during the inference stage without the use of manually-determined audio.
Considering that the style and semantic information of sentences are closely related to each other, the PLMs, such as BERT \cite{bert}, have been used to provide richer semantic information for expressive speech synthesis \cite{berttacotron, bertemb, fang2019towards}. 
In these works, the semantic representations derived from pre-trained BERT are added to the acoustic model as additional inputs according to attention mechanism or forced-alignment technology.
Several previous works \cite{wsv, liu22m_interspeech,ren2022prosospeech, tan2020fine, du2021rich} attempt to model fine-grained style from text in order to control local prosodic variations in speech.
The word-level style variation (WSV) \cite{wsv} designs an unsupervisedly-learned fine-grained representations to describe local style properties.
Multiple studies combine word-level style modeling with the non-autoregressive acoustic model to improve the efficiency of speech synthesis \cite{liu22m_interspeech, ren2022prosospeech}, and other studies further model the finer-grained style, such as at phoneme level \cite{tan2020fine, du2021rich}.

Our work is distinct from the above in two aspects: 
(i) previous models only focus on mono-scale style information while our work takes into consideration style information at difference levels beyond the sentence level;
(ii) other models utilize the semantic information of only the current sentence to improve pronunciation and expressiveness of the synthesized speech,
while we make use of the context information containing multiple adjacent sentences.
To our best knowledge, there is currently no research conducting multi-scale style modeling from the context for expressive speech synthesis.

\subsection{Context-aware Style Modeling}
Recently, some efforts have been made to consider a wider range of contextual information to improve the performance of expressive speech synthesis.
\cite{nakata11audiobook} obtains context-aware text representations of the current sentence by feeding both the neighboring and current sentences to BERT.
Moreover, some recent researches \cite{crossutterance, li2022cross} have further utilized cross-utterance information obtained from neighboring sentences to improve the prosody.
Apart from the contextual text information, it has been reported that contextual speech information can also lead to improvement of expressiveness \cite{oplustil2020using, gallegos2021comparing}.
In addition to synthesize speech at the sentence-level, \cite{xue2022paratts} has also implemented paragraph-based speech synthesis by inputting phoneme sequence of the entire paragraph. 

Above methods use context information as additional inputs to implicitly learn the prosody or style of speech.
Different from this implicit way, our proposed method utilizes a multi-scale style extractor to extract style representations at the three levels from ground-truth speech and explicitly guide the style prediction.
In addition to the contextual semantic information, the hierarchical structural relationships in context are also considered in our method.
Then, with the help of two style predictors, style embeddings of each level are generated for single sentence or whole paragraph.
Moreover, compared with ParaTTS \cite{xue2022paratts} that considers phoneme sequence only, our proposed MSStyleTTS further considers the semantic and structural information in the context and models the style at different levels in the paragraph in an explicit way.


\section{Proposed Method} \label{sec:model}
\begin{figure*}[!tb]
	\centering
	\includegraphics[width=0.9\linewidth]{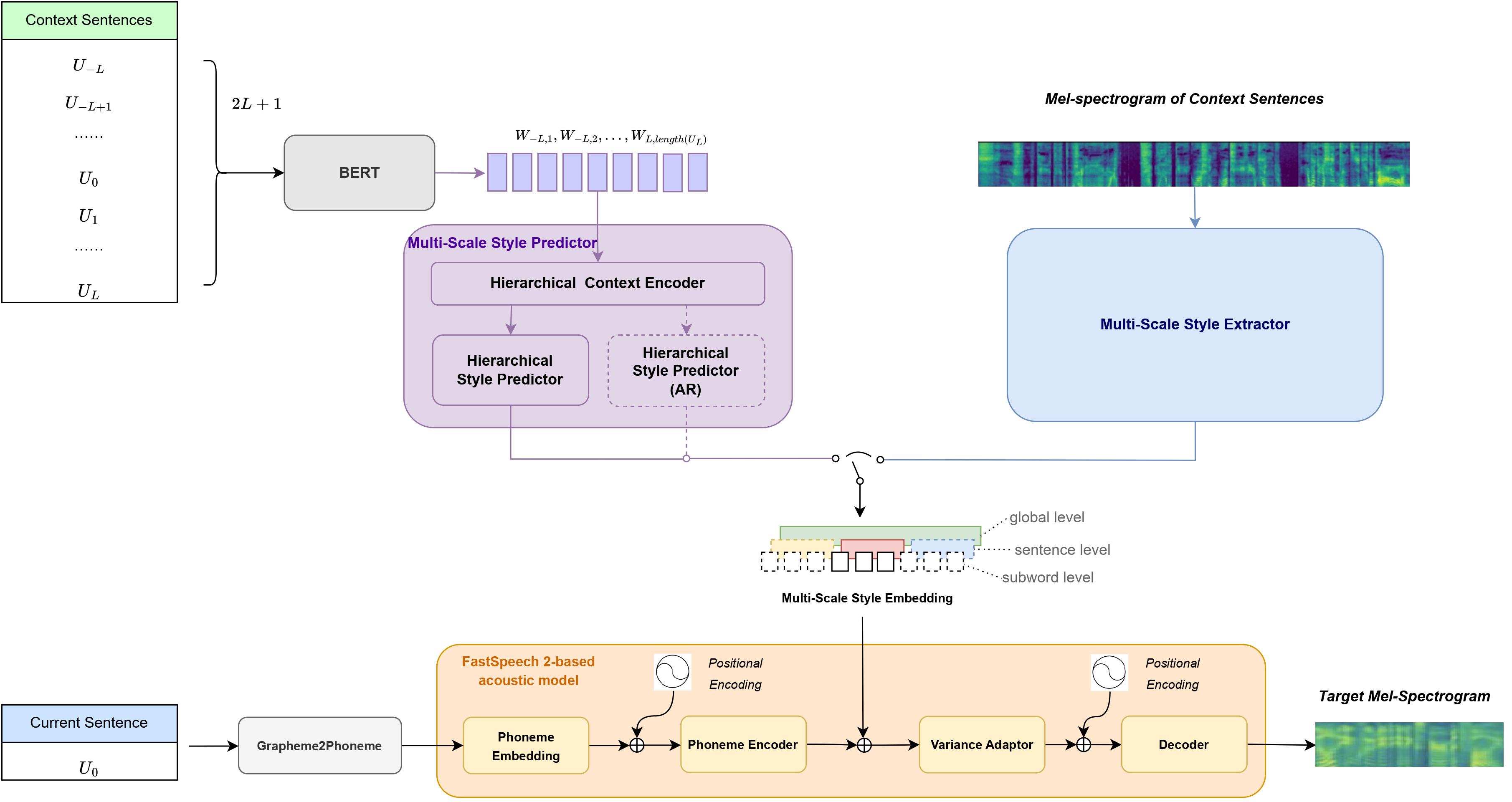}
	\caption{MSStyleTTS model architecture. The blue boxes represent the multi-scale style extraction procedure, purple boxes represent multi-scale style prediction procedure, and yellow boxes are the TTS backbone.}
	\label{fig:architecture}
\end{figure*}
\subsection{Model Architecture} \label{sec:modelarchitecture}
The framework of MSStyleTTS is presented in Fig.\ref{fig:architecture}.
It consists of a modified FastSpeech 2 \cite{fastspeech2} as acoustic model 
to generate mel-spectrogram from phoneme sequence,
a multi-scale style extractor and a multi-scale style predictor to learn the style information at global level, sentence level, and subword level.
The extractor is utilized to respectively extract the three levels of style embedding from the corresponding ranges of mel-spectrogram.
The predictor is utilized to model the style embedding at above three levels from a wider range of context beyond a sentence.
The multi-scale style embedding obtained from either the extractor or the predictor is then added to the output of the phoneme encoder and fed to the variance adaptor for predicting speech variations more accurately and for generating speech with expressive style.
The details of each component are as follows.

\subsubsection{Multi-Scale Style Extractor} \label{sec:modelextractor}
\begin{figure}[!tb]
	\centering
	\includegraphics[width=0.8\linewidth]{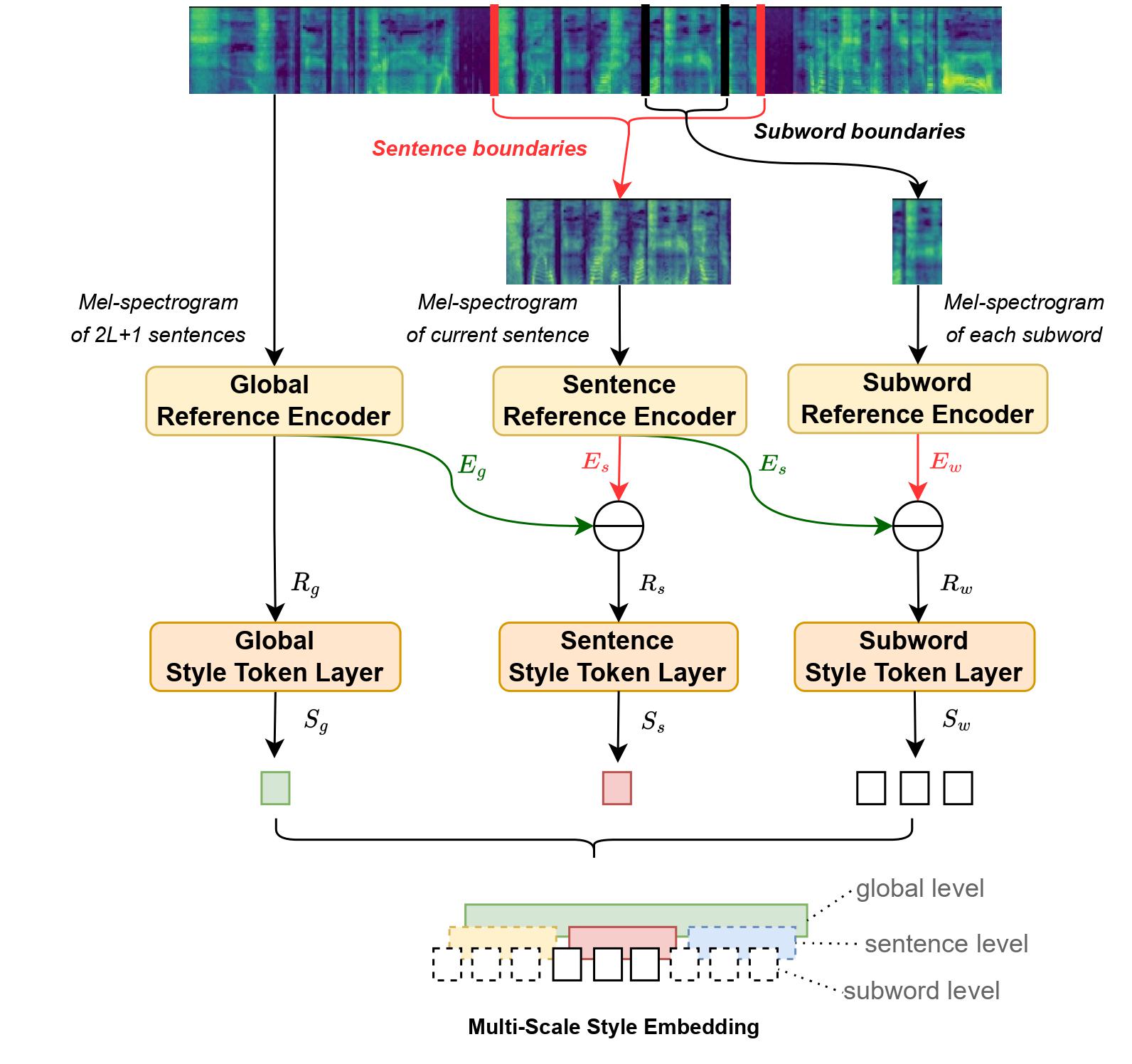}
	\caption{Architecture of the multi-scale style extractor.}
	\label{fig:extractor}
\end{figure}
The multi-scale style extractor is specifically designed to derive three levels of style embedding from a wider range of mel-spectrogram rather than a sentence.
As shown in Fig.\ref{fig:extractor}, the extractor is made up of three reference encoders and three style token layers corresponding to the global-level, sentence-level, and subword-level.
The structure and hyperparameters of each of these modules remain the same as those of the original GST \cite{gst} model.

The reference encoder contains 6 convolution layers followed by a gated recurrent unit (GRU) \cite{gru} layer to extract high-level representation from the mel-spectrogram. 
In addition to the current sentence, the mel-spectrograms of $L$ past sentences and $L$ future sentences are also considered to obtain the global-level style in our method.
The ground-truth mel-spectrograms of all $2L+1$ sentences are concatenated before passing through the global reference encoder.
The output of this encoder is denoted as global-level reference embedding $E_g$.
The mel-spectrogram of the current sentence is then passed to the sentence reference encoder, whose output is denoted as sentence-level reference embedding $E_s$.
Meanwhile, the mel-spectrogram corresponding to the current sentence is split by the subword boundaries and fed to the subword reference encoder.
The subword reference encoder extracts the subword-level reference embedding $E_w$ for each subword from the corresponding mel-spectrogram.
In our practice, the subword boundaries are derived from the subword-to-phoneme alignments and forced-alignment phoneme boundaries.

The information from the extracted reference embeddings of different levels may be overlapped since the wider range mel-spectrogram covers the smaller range mel-spectrogram.
More specifically, the finer-grained reference embedding $E_w$ may contain redundant style information that has already been covered in the coarser-grained reference embedding $E_s$, and likewise for $E_s$ and $E_g$. 
To address this issue, instead of using the extracted reference embeddings directly, the residual embeddings are proposed to represent effective style variations at different levels.
The residual embeddings at different levels can be obtained from the following functions:
\begin{align}
    R_g &= E_g, \\
    R_s &= E_s-E_g, \\
    R_w &= E_w-E_s,
\end{align}
where $R_g$, $R_s$, and $R_w$ are the residual embeddings corresponding to global-level, sentence-level, and subword-level, respectively.
Then the residual embedding of each level goes through the style token layer of the corresponding level and is decomposed into a set of style tokens.
These tokens are trained to learn the stylistic or prosodic information at different levels and help to predict the style.
Here, we regard the output of the style token layer of each level as the style embedding of the corresponding level, denoted as the global-level style embedding $S_g$, sentence-level style embedding $S_s$, and subword-level style embedding $S_w$, respectively.

\subsubsection{Multi-Scale Style Predictor} \label{sec:modelpredictor}
\begin{figure}[!tb]
	\centering
	\includegraphics[width=1\linewidth]{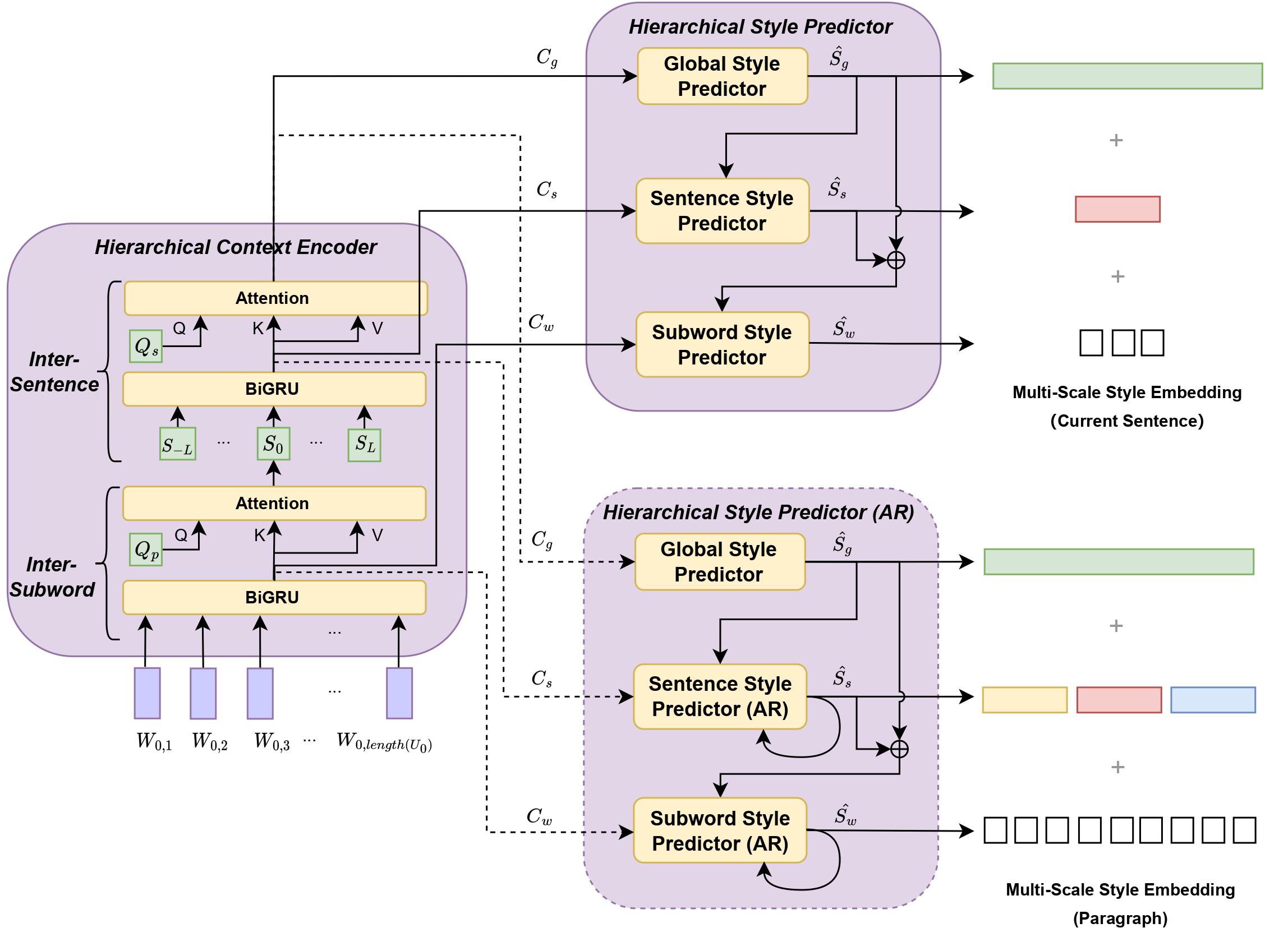}
	\caption{Architecture of the multi-scale style predictor.}
	\label{fig:predictor}
\end{figure}
During inference, instead of extracting multi-scale style embedding from manually-selected reference audio, we propose a multi-scale style predictor to predict the style embedding at different levels from a fixed size sliding window containing past, current and future sentences.
The architecture of the multi-scale style predictor is illustrated in Fig.\ref{fig:predictor}, which comprises a hierarchical context encoder and two optional hierarchical style predictors.

To derive better text representation as predictor input, we introduce a pre-trained BERT to further encode the raw text sequence for Chinese.
Together with the current sentence, the predictor also take into account past and future $L$ sentences, which is consistent with the extractor.
Let $U_0$ denote the text sequence of the current sentence.
$U_{-L}$, $U_{1-L}$, ..., $U_{-1}$ and $U_{1}$, $U_{2}$, ..., $U_{L}$ are the text sequences of past and future sentences, respectively.
The text sequences of all $2L+1$ sentences are concatenated to form a longer sequence $U$, which is then fed to BERT to obtain a subword-level semantic embedding sequence. 
It is calculated as follows:
\begin{gather}
    U = Concat(U_{-L},U_{1-L},\cdots,U_{L}), \\
    W_{-L,1},W_{-L,2},\cdots,W_{L,length(U_L)} = BERT(U),
\end{gather}
where $Concat(\bigcdot)$ is the concatenation operation, $length(U_i)$ is the number of subwords in $U_i$, and $W_{i,j}$ is the 768-dim subword-level semantic embedding corresponding to the $j$th subword of sentence $U_i$ in BERT outputs.

Inspired by \cite{han}, we design a hierarchical context encoder to derive three levels of context information.
The hierarchical context encoder has two layers of attention modules, i.e., the inter-subword module and inter-sentence module, to explicitly exploit the structural relationship of the context.
These two modules share the same architecture consisting of a bidirectional GRU \cite{gru} and a scaled dot-product attention \cite{vaswani2017attention}.
The inter-subword module is utilized to derive the sentence-level representation based on the semantic of each subword and the inter-subword relationships in a sentence.
Specifically, for each sentence, the semantic embedding sequence derived from BERT is fed into the bidirectional GRU to consider the temporal relationship, and the output is regarded as subword context embedding $C_w$.
Since that not all subwords contribute equally to the meaning of their sentence, we introduce the attention module to derive the weight of each subword and then aggregate them into a sentence-level embedding. 
Likewise, the inter-sentence module is utilized to derive the global-level representation based on each sentence-level embedding and inter-sentence relationships in the context.
We denote the output of bidirectional GRU in the inter-sentence module as sentence context embedding $C_s$ and the output of attention module in the same module as global context embedding $C_g$.
It is noteworthy that this hierarchical structure encodes inter-subword and inter-sentence relationships in a wide range of context, which contributes to be a hierarchical information aware model.

Considering the top-down hierarchical structure of speech style, we design a hierarchical style predictor to infer the style embedding at the three levels from corresponding context information in a top-down manner.
It consists of three style predictors with residual connections, where each style predictor consists of a linear layer activated by Tanh function.
The coarser-grained style embedding that is closer to the global-level is predicted by the corresponding level of style predictor, and then fed into the finer-grained style predictor as the conditional input, which is symmetrical with the residual strategy in the multi-scale style extractor.
In this way, the style embedding at three levels are sequentially generated from the three style predictors by considering both context information and stylistic information.
This can be formulated as:
\begin{align}
    \hat{S_g} &= f_g(C_g), \\
    \hat{S_s} &= f_s(C_s,\hat{S_g}), \\
    \hat{S_w} &= f_w(C_w,\hat{S_g}+\hat{S_s}),
\end{align}
where $f_g$, $f_s$, and $f_w$ are the style predictors corresponding to global-level, sentence-level, and subword-level, respectively. $\hat{S_g}$, $\hat{S_s}$, and $\hat{S_w}$ are the predicted style embeddings corresponding to global-level, sentence-level, and subword-level. 
In this way, the style embeddings at above three levels are predicted to reconstruct the multi-scale style in human recordings by considering different levels of context information.

To synthesize speech directly for paragraph-level text input, we propose an optional extension of the hierarchical style predictor. 
Since the style coherence among sentences is equally important for paragraph-level speech in addition to expressiveness, this predictor utilizes previous style embedding in an autoregressive manner to help the prediction of the current style embedding, denoted as AR.
Different from the hierarchical style predictor, the hierarchical style predictor (AR) utilizes GRU-based sentence style predictor (AR) and subword style predictor (AR) to sequentially predict the style of sentences in the whole paragraph.
In the sentence style predictor (AR), the sentence-level style embedding of each sentence is predicted autoregressively, using the corresponding sentence context embedding and the predicted global-level style embedding as the conditions.
Similarly, the subword-level style embedding of each subword is predicted autoregressively in subword style predictor (AR), which uses corresponding subword context embedding and the coarser-grained style embedding corresponding to its sentence as the conditions.
Finally, the style embeddings corresponding to each sentence are fed into the acoustic model to generate the speech of this sentence.
\subsubsection{Acoustic Model}
As illustrate in Fig.\ref{fig:architecture}, our proposed MSStyleTTS adopts a modified FastSpeech 2 \cite{fastspeech2} as the acoustic model.
First, the extracted or predicted multi-scale style embeddings are replicated to the phoneme-level using subword-to-phoneme alignment.
After replication, the embeddings of the phonemes are added to the outputs of the phoneme encoder and fed into the variance adaptor to predict variations more accurately. 
Moreover, the duration predictor and length regulator are moved to the back of the variance adaptor.
This indicates that the variance adaptor predicts phoneme-level variations rather than frame-level, which has been demonstrated to enhance the naturalness of synthesized speech \cite{fastpitch}.
Here, FastSpeech 2 is only utilized to combine our proposed predictor and extractor to generate acoustic features.
Our proposed method can be easily extended to
other TTS models, such as Tacotron2 \cite{tacotron2}.

\subsection{Model Training} \label{sec:modeltraining}
In general, it is difficult for the predictor to implicitly predict different levels of styles from the context under the situation of limited TTS data.
The parameters of MSStyleTTS are trained in three steps using the knowledge distillation strategy, in order to encourage the model learn style representation better.

At first, the extractor is jointly trained with the acoustic model to learn the latent style representations at different levels in an unsupervised manner.
For each training utterance, the phoneme sequence of the current sentence and a wider range of mel-spectrogram containing adjacent sentences are sent into the acoustic model and the extractor, respectively.
We conduct the same training criteria as FastSpeech 2, which includes a mel-spectrogram loss and the mean square error (MSE) loss of pitch, energy, and duration.
Moreover, we train the reference encoders and style token layers corresponding to global-level, sentence-level, and subword-level sequentially and independently, in order to avoid the leaning of style representations at different levels interfering with one another. 
That is, while training any of these three level modules, the others are frozen, so that the reference encoder and style token layer of the current level are trained without disturbance. 

Second, we transfer the knowledge from the extractor to the predictor by leveraging knowledge distillation strategy.
Thus, the extracted style embeddings for each level are used as the target outputs to guide the predictor's training.
For each training utterance, its context, which contains current and adjacent sentences, is employed to predict the styles at different levels by the predictor.
The predictor is trained by minimizing the sum of the MSE losses between the extracted and predicted style embeddings of the three levels.

In the end, the acoustic model is jointly trained with the predictor at a lower learning rate by considering both the losses of the predicted mel-spectrogram and style to improve the naturalness of synthesized speech further.
In particular, for the hierarchical style predictor (AR), we generate the mel-spectrogram for each sentence in the paragraph and calculate the corresponding losses of mel-spectrogram and style.

\subsection{Inference}
During the inference for the current sentence, the proposed method automatically derives the multi-scale style embedding from the context and generates mel-spectrograms based on the phoneme sequence of the current sentence with the help of the style embedding.
During the inference for the paragraph-based synthesis, we prefer to use the hierarchical style predictor (AR) which directly predicts the multi-scale style embedding for the entire paragraph.
Then we send the multi-scale style embedding and phoneme sequence corresponding to each sentence into the acoustic model to generate mel-spectrograms of each sentence, and then combine them in order.
The predicted mel-spectrogram goes through the HiFi-GAN \cite{kong2020hifi} vocoder to obtain the speech waveform.
In this way, expressive speech can be synthesized without the dependency on additional inputs besides text.
\section{Experiments} \label{sec:exp}
\subsection{Experimental Setup}
In our experiments, an internal single-speaker audiobook corpus is adopted to train and evaluate all the models\footnote{Implemented based on: \href{https://github.com/ming024/FastSpeech2}{https://github.com/ming024/FastSpeech2}\label{fn_fs2}}.
This corpus contains around 87 chapters of a fiction novel recorded by a professional Chinese male speaker, about 30 hours in total.
This audiobook corpus is suitable for expressive speech synthesis, as the styles vary between utterances, and the prosody characteristics fluctuate significantly within each utterance, such as pitch, energy, and speed.
The corpus is divided into 14,558 utterances, of which 200 are used for testing, 456 for validation, and the remainder for training.
To further test the performance of the models outside the training corpus domain, we select 30 sentences with its context from other different audiobooks as the evaluation dataset for the out-of-domain experiments.

All the text sequence is transformed to the phoneme sequence and provided as input to the acoustic model.
The 80-dimensional mel-spectrograms are extracted from the raw waveform with frame size 1,200, hop size 240, and sampling rate 24kHz, and are used as the target output.
Silences at the beginning and end of each utterance are trimmed.
We force-align the audio and phoneme sequence by an automatic speech recognition tool to obtain phoneme boundaries and durations.
Meanwhile, we use the Python library of PyWORLD\footnote{\href{https://github.com/JeremyCCHsu/Python-Wrapper-for-World-Vocoder}{https://github.com/JeremyCCHsu/Python-Wrapper-for-World-Vocoder}} to extract the frame-level pitch and energy values, and then derive the pitch and energy values at phone-level by averaging frame-level values based on the forced-aligned phoneme boundaries.

In our implementation, the sentences considered in the extractor and predictor are made up of the current sentence, its two past sentences, and its two future sentences.
We conduct an open-source pre-trained Chinese BERT-base model released by Google\footnote{\href{https://github.com/google-research/bert}{https://github.com/google-research/bert}} to extract the semantic information.
Its parameters are fixed when training the predictor.
The model is trained using a single NVIDIA V100 GPU with a batch size of 16 for 220k steps.
For MSStyleTTS, 180k iterations are required for the first train step, with 60k for each of the global-level, sentence-level, and subword-level modules, then 20k for the second train step and 20k for the final train step.
We utilize Adam optimizer\cite{kingma2014adam} with $\beta_1=0.9$, $\beta_2=0.98$, $\epsilon=10^{-9}$ and employ the warm-up strategy before 4,000 iterations.
Moreover, a well-trained HiFi-GAN \cite{kong2020hifi} model is utilized as the vocoder to transform the mel-spectrogram into the waveform, which is pre-trained on a large standard corpus and is adapted with 200k iterations on our audiobook corpus.

\subsection{Baseline Methods}
In order to test the performance of our proposed multi-scale model, we choose three baselines for comparison.
All baselines and proposed models only use text input at the inference stage, and share the same input context sequence and phoneme sequences.
The details of the three baseline models are described as follows:

\textbf{FastSpeech 2}
An open-source implementation\textsuperscript{\ref{fn_fs2}} of FastSpeech 2 \cite{fastspeech2}.
The model architecture is consistent with that of the acoustic model in our MSStyleTTS.

\textbf{WSV*}
To build this baseline model, the Word-Level Style Variations (WSVs)\cite{wsv} approach with several modifications is followed, which is referred as WSV*.
To ensure a fair comparison, FastSpeech 2 is employed as the acoustic model in our implementation, rather than the Tacotron 2 \cite{tacotron2} in the original version of WSV \cite{wsv}.
Moreover, we input the same context as MSStyleTTS into BERT, and then utilize an extra bidirectional GRU to consider the context information.
Thus it is also a context-aware method and models style at the word level.

\textbf{HCE} 
We conduct the Hierarchical Context Encoder (HCE) \cite{proposed} model which models the style at the global level from the context.
For a fair comparison, instead of XLNet \cite{xlnet} which is used in the original version of HCE, BERT has been used to obtain semantic information. 
The structure and hyperparameters of the pre-trained BERT model, acoustic model, and hierarchical context encoder are the same as those in MSStyleTTS.

\subsection{Subjective Evaluation}
\begin{table*}[h]
  \caption{The MOS test scores of different models with 95\% confidence intervals.}
  \label{tab:mos}
  \centering
  \begin{tabular}{c|ccc} 
    \hline
    \textbf{Models} & \multicolumn{2}{c}{\textbf{S-MOS}} & \textbf{M-MOS}  \\
    \textbf{} & \textbf{In-domain} & \textbf{Out-of-domain} & \textbf{}  \\
    \hline
    Ground Truth & $4.665\pm0.074$ & - & $4.560\pm0.071$ ~~~               \\
    FastSpeech 2 & $3.573\pm0.094$& $3.569\pm0.057$& $3.669\pm0.072$  ~~~  \\
    WSV* & $3.681\pm0.080$& $3.579\pm0.059$& $3.664\pm0.068$ ~~~ \\
    HCE & $3.785\pm0.084$& $3.681\pm0.065$& $3.682\pm0.075$ ~~~  \\
    \hline
    MSStyleTTS & $\mathbf{4.058\pm0.074}$ & $\mathbf{3.997\pm0.061}$ & $3.758\pm0.069$~~~ \\
    MSStyleTTS (AR) & - & - & $\mathbf{3.867\pm0.072}$~~~ \\
    \hline
  \end{tabular}
\end{table*}
\begin{figure}[!tb]
	\centering
	\includegraphics[width=1\linewidth]{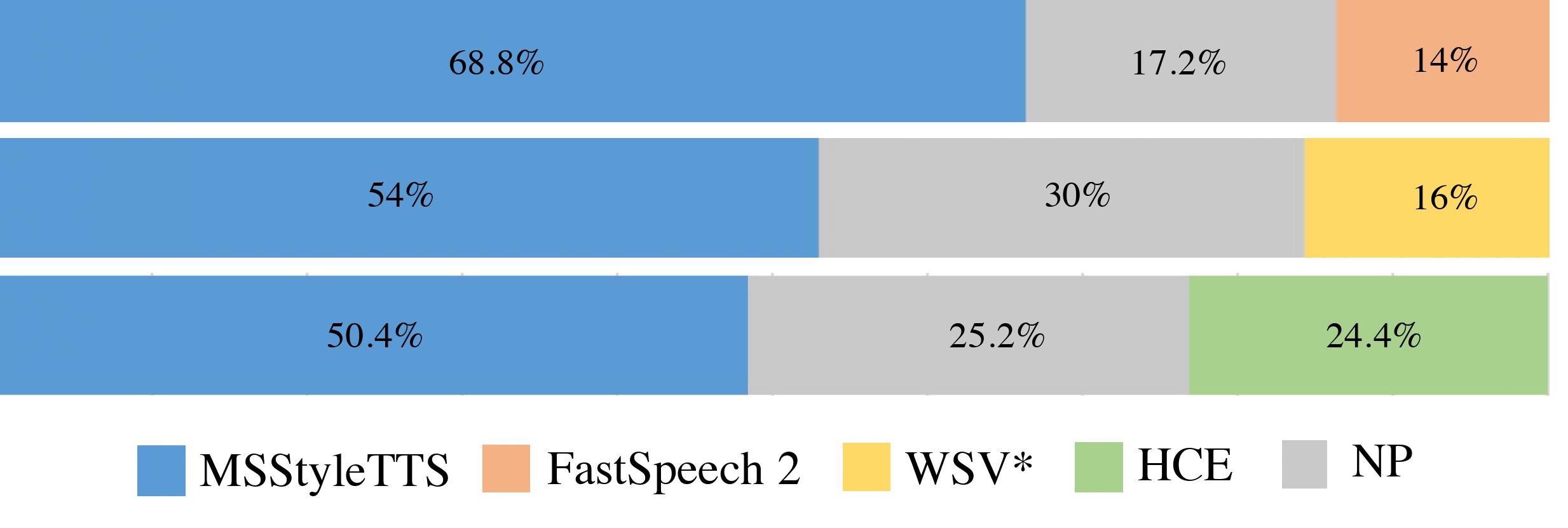}
	\caption{The ABX preference test results on naturalness and expressiveness among the proposed MSStyleTTS and three baseline models. NP means ``no preference".}
	\label{fig:abx}
\end{figure}
Two mean opinion score (MOS) tests are applied to evaluate the naturalness and expressiveness of the synthesized speech: 1) Single-sentence MOS (S-MOS): evaluate the synthesized speech of single sentence; 2) Multi-sentence MOS (M-MOS): evaluate the entire long-form speech which combines the synthesized speeches of multiple sentences in order.
A group of 25 listening subjects who are native Mandarin speakers are recruited to rate the generated speeches on a scale from 1 to 5 with 1 point interval.
They also score the audio reconstructed from the ground truth mel-spectrogram by HiFiGAN vocoder.
Additionally, we conduct the ABX preference test between our proposed MSStyleTTS and each of the three baselines to further analyze the perceived quality of the synthesized speech.
The same 25 listeners are asked to give their preferences between a pair of speeches synthesized by different models.
\subsubsection{S-MOS results}
The results are shown in Table \ref{tab:mos}.
For the in-domain dataset, it is observed that FastSpeech 2 has achieved the lowest MOS, with a significant gap compared to Ground Truth, suggesting that it is difficult to directly predict the rich and complex prosodic variations only from phonetic descriptions.
By considering context information, the other three models (WSV*, HCE, and MSStyleTTS) all perform better.
Our MSStyleTTS achieves the highest score of $4.058$, exceeding WSV* that just considers the fine-grained style representation by $0.377$ and HCE that narrowly considers the coarse-grained style representation by $0.273$.
This demonstrates that while each mono-scale baseline has different degrees of improvement in the expressiveness of the synthesized speech, the overall performance can be significant enhanced by merging them into a comprehensive multi-scale style model. 
For the out-of-domain dataset, we observe a decrease in scores for each model compared with the in-domain dataset, and MSStyleTTS still receives the highest score.
It indicates that our proposed model successfully learns to model different levels of style representations with robustness and generalization.

\subsubsection{M-MOS results}
The last column in Table \ref{tab:mos} shows the results of M-MOS.
We observe that speech synthesized by FastSpeech 2 has relatively smooth style transitions among sentences but lacks of expressiveness, while the speech synthesized by WSV* and HCE has rich expressiveness but varies greatly from one sentence to the next, both of which result in poor performance in the paragraph-based synthesis.
Benefiting from modelling the style representation beyond the sentence level, our proposed model achieves a higher score of $3.758$. 
Moreover, with the help of autoregressive style predictor, the M-MOS score of the proposed MSStyleTTS (AR) gains further improvement compared to the MSStyleTTS.
This demonstrates that in paragraph-based synthesis, the style coherence of the synthesized speech is important for the perceived quality.

\subsubsection{ABX test results}
The preference test results are presented in Fig.\ref{fig:abx}, which demonstrate that our proposed MSStyleTTS model significantly outperforms all the baselines.
With the additional context semantic information, MSStyleTTS gets an extra preference ($54.8\%$) over FastSpeech 2.
In addition, our MSStyleTTS is $38\%$ more preferred than the WSV*, indicating that modeling the coarse-grained style can synthesize speech with richer expressiveness.
Compared with HCE, the advantage of MSStyleTTS is that fine-grained style representation is able to control the local style characteristics, such as intonation and stress pattern, leading to a higher preference of $26\%$ than HCE.
\begin{table}[!tb]
  \caption{Objective evaluation results for MCD, F0 RMSE, Energy RMSE and Duration MSE of different models on the test set.}
  \label{tab:objective}
  \centering
  \begin{tabular}{ccccc} 
    \hline
     &\textbf{FastSpeech 2} & \textbf{WSV*} & \textbf{HCE} & \textbf{MSStyleTTS}\\
    \hline
    MCD  & $5.066$ & $5.062$ & $5.031$ & \textbf{4.979}~~  \\
    F0 RMSE & $65.266$ & $64.807$ & $63.683$& \textbf{62.544}~~ \\
    Energy RMSE  & $5.162$ & $5.221$ & $5.045$ & \textbf{4.926}~~  \\
    Duration MSE  & $0.218$ & $0.205$ & $0.209$ & \textbf{0.201}~~  \\
    \hline
  \end{tabular}
\end{table}
\subsection{Objective Evaluation}
To measure the prosody and naturalness of synthesized speech objectively, we calculate mel-cepstrum distortion (MCD), the root mean square error (RMSE) of F0 and energy, and the MSE of duration as the metrics of objective evaluation following \cite{fastspeech2, xue2022paratts}.
Since the lengths of the predicted and ground truth mel-spectrograms may be different, we first apply dynamic time warping (DTW) to derive the alignment relationships between the two mel-spectrograms.
Then, we compute the minimum MCD by aligning the two mel-spectrograms.
Following the alignment relationships, the F0 and energy sequences extracted from the synthesized speech are also aligned towards ground truth.
For the duration, we directly calculate the MSE between the predicted phoneme duration and the ground truth one.

The objective evaluation results are presented in Table \ref{tab:objective}.
The model with fine-grained style modeling (WSV*) or coarse-grained style modeling (HCE) outperforms FastSpeech 2 in most metrics, and our MSStyleTTS is the best.
From the results, it is observed that even though the fine-grained style is more closely associated with the prosody, WSV* does not show better performance than HCE in terms of the RMSE of F0 and energy.
Possible reason is that the hierarchical structure of context is ignored.
Better performance is achieved by MSStyleTTS with the multi-scale style modeling framework, demonstrating the advantages of modeling style at different levels.
It restores more accurate prosody characteristics such as pitch, energy, and duration for expressive speech synthesis.

\begin{figure}[!tb]
	\centering
	\includegraphics[width=1.0\linewidth]{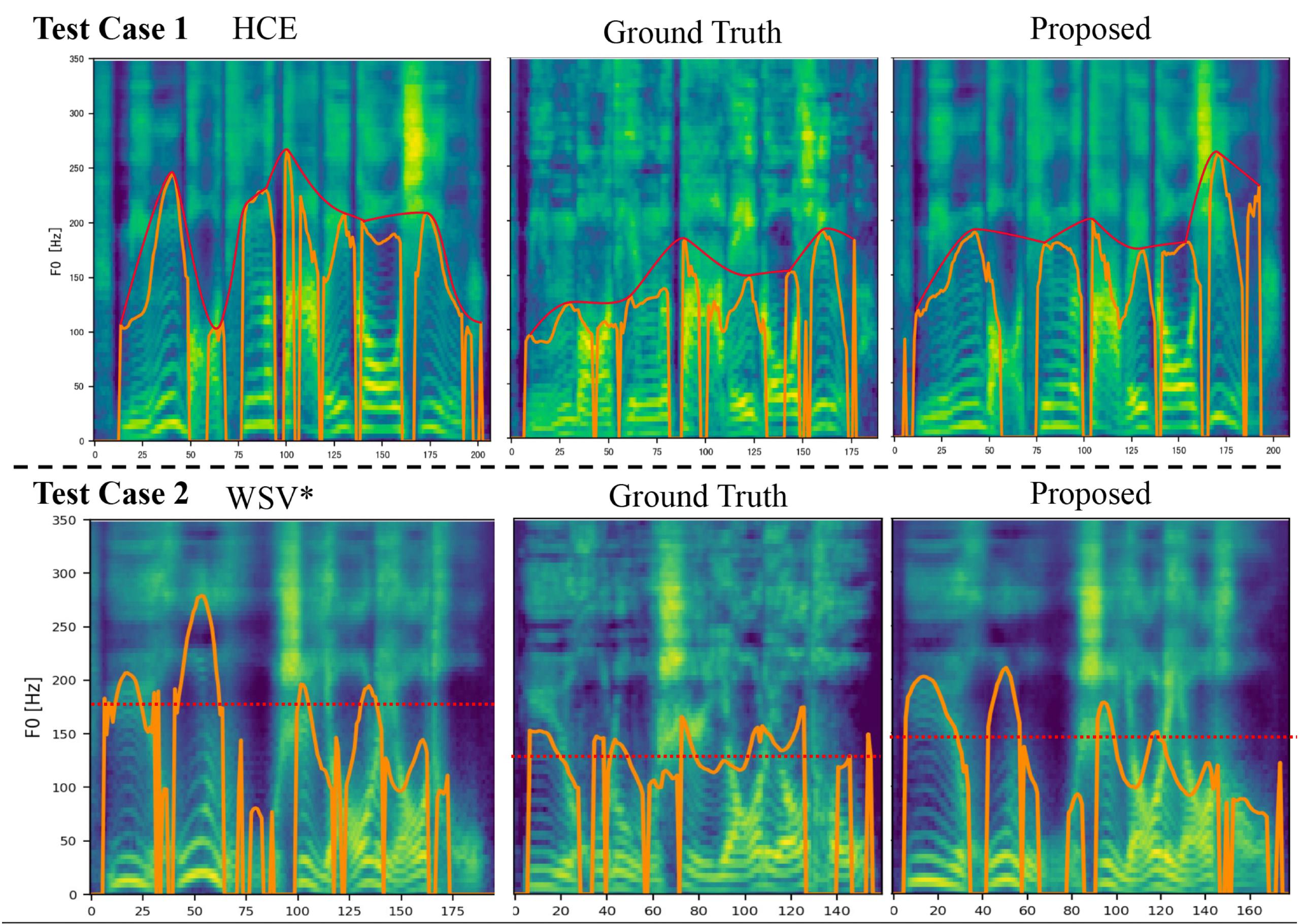}
	\caption{Mel-spectrograms and pitch contours of ground truth speech and speech synthesized by different models for two test cases. The red line in test case 1 represents the intonation trend, and the red dashed line in test case 2 represents the average pitch value. The text transcription of test case 1 is ``míng bǎo tīng le dà wéi gāo xìng a" (English translation: ``Ming Bao was very happy to hear that"), and the text transcription of test case 2 is ``yù hú lǔ shì yáo le yáo tóu" (English translation: ``Yuhulu shakes her head").}
	\label{fig:casestudy}
\end{figure}

\subsection{Case Studies}
To further explore the impact of the multi-scale style modeling framework on the expressiveness and prosody of synthesized speech, two case studies are conducted to compare our MSStyleTTS with two mono-scale baselines, respectively.
The ground truth speeches are also provided as references.
Fig.\ref{fig:casestudy} shows the mel-spectrograms and pitch contours of ground truth speeches and speeches synthesized by different models for two test utterances.

It is observed that the speech synthesized by HCE contains larger pitch fluctuation than our MSStyleTTS.
Nevertheless, due to the absence of fine-grained style, it is difficult to control the local style variations of generated speech, resulting in a significant difference in the intonation trend compared with ground truth.
The intonation trend of the speech synthesized by WSV* is similar to MSStyleTTS but has a higher overall pitch value that is inconsistent with ground truth.
Compared with these two mono-scale baselines, the proposed multi-scale MSStyleTTS generates speech that is more similar to ground truth in terms of both the overall pitch value and local style variations, such as the intonation trend and stress patterns.

\begin{table*}[t]
  \caption{Objective evaluation results of the proposed extractor and the extractor without residual strategy on the test set. Here, ``extracted style" and ``predicted style" stand for synthesized speech with the multi-scale style embedding provided by the extractor and predictor respectively. ``-residual strategy" stand for removing the residual embeddings used in the extractor.}
  \label{tab:objextractor}
  \centering
  \begin{tabular}{lcccc} 
    \hline
     \textbf{Models} &\textbf{MCD} & \textbf{F0 RMSE} & \textbf{Energy RMSE} & \textbf{Duration MSE}\\
    \hline
    Proposed (extracted style)  & $4.710$ & $44.457$ & $3.237$ & $0.098$~~  \\
    \quad -residual strategy (extracted style) & $4.723$ & $44.416$ & $3.116$& $0.097$~~ \\
    Proposed (predicted style) & $4.979$ & $62.544$ & $4.926$ & $0.201$~~  \\
    \quad -residual strategy (predicted style) & $5.013$ & $62.956$ & $4.993$ & $0.203$~~  \\
    \hline
  \end{tabular}
\end{table*}

\begin{table}[h]
  \caption{Results of the CMOS test between MSStyleTTS and the model without using residuals to represent style variations.}
  \label{tab:cmosextractor}
  \centering
  \begin{tabular}{l|c} 
    \hline
    \textbf{Models} &\textbf{CMOS} \\
    \hline
    MSStyleTTS & $0$ ~~~ \\
    \quad -residual strategy & $-0.516$ ~~~ \\
    \hline
  \end{tabular}
\end{table}

\begin{table}[h]
  \caption{Results of the CMOS test between MSStyleTTS and the model without using the knowledge distillation strategy to train the predictor.}
  \label{tab:cmostrain}
  \centering
  \begin{tabular}{l|cc} 
    \hline
    \textbf{Models} &\multicolumn{2}{c}{\textbf{CMOS}} \\
    \textbf{} & \textbf{In-domain} & \textbf{Out-of-domain} \\
    \hline
    MSStyleTTS & $0$ & $0$ ~~~ \\
    \quad -knowledge distillation & $-0.620$ & $-0.469$ ~~~ \\
    \hline
  \end{tabular}
\end{table}
\subsection{Ablation Studies}
We conduct ablation studies to demonstrate the effectiveness of several techniques used in our proposed MSStyleTTS, including using knowledge distillation strategy to train the predictor,  using residuals to represent style variations for style extraction, utilizing context information in the predictor, using multi-scale style predictor for style prediction, and designing multi-scale style representations.

\subsubsection{The Effect of Using Knowledge Distillation Strategy to Train the Predictor}
As introduced in Section \ref{sec:modeltraining}, our proposed model is trained with a knowledge distillation strategy in three steps.
In this part, we conduct ablation studies by removing this strategy.
That is, the multi-scale style predictor is jointly trained with the acoustic model directly.
Table \ref{tab:cmostrain} presents the results.
We can see that using knowledge distillation strategy can gain signiﬁcant improvements in both in-domain and out-of-domain datasets.
It demonstrates that the predictor can learn better style representations of different levels by utilizing a pre-trained extractor as teacher model.

\subsubsection{The Effect of Using Residuals to Represent Style Variations}
As shown in Fig.\ref{fig:extractor}, we represent the style variations at different levels as residuals.
To verify the effectiveness of this representation method, we build a new extractor without residual strategy.
Firstly, we synthesize speeches with style embeddings extracted from ground-truth speech by two extractors, which is also regarded as the upper bound performance of our proposed technique.
The objective evaluation results are presented in the first two rows of Table \ref{tab:objextractor}.
As can be seen, the style embeddings extracted by two extractors have comparable performance, indicating that using residual strategy does not affect the power of style extraction.

Considering the purpose of extracting style embedding is to guide the training of the predictor rather than style transfer, we further trained two style predictors by utilizing these two style embeddings as target.
The last two rows in Table \ref{tab:objextractor} show the objective evaluation results of the generated speeches based on the predicted style embedding.
We can see that removing the residual strategy performs worse in terms of all evaluation metrics, indicating that using residuals to represent style variations is beneficial for style prediction.

The comparison mean opinion score (CMOS) test further compares perceived quality of the generated speeches.
The listening subjects are asked to rate the comparative degree between two compared models on a scale from -3 to 3 with 1 point interval, where lower rate indicates that the first model performs better than the second model and higher rate indicates that the second model performs better than the first model.
As shown in Table \ref{tab:cmosextractor}, the neglect of residual strategy results in $-0.516$ CMOS.
These results indicate that our proposed style extraction approach can effectively obtain multi-scale style information from the speech and facilitate the subsequent prediction task by reducing the interference or overlap between different levels of styles.

\subsubsection{Comparisons of Utilizing Different Ranges of Context Information in Predictor}
\begin{table}[t]
  \caption{Objective evaluation results of the proposed MSStyleTTS on the test set when using different ranges of context in predictor. Here, ``L" stands for the number of sentences considered in both the past and future context. ``L=0" means only considering the text of the current sentence.} 
  \label{tab:objcontext}
  \centering
  \begin{tabular}{ccccc} 
    \hline
    L &\textbf{MCD} & \textbf{F0 RMSE} & \textbf{Energy RMSE} & \textbf{Duration MSE}\\
    \hline
    0  & $5.041$ & $64.410$ & $5.155$ & $0.207$~~  \\
    1 & $5.016$ & $62.704$ & $5.108$& $0.207$~~ \\
    2 & \textbf{4.979} & \textbf{62.544} & \textbf{4.926} & \textbf{0.201}~~  \\
    3 & $5.048$ & $63.130$ & $5.082$ & $0.206$~~  \\
    4 & $5.121$ & $63.277$ & $5.143$ & $0.210$~~  \\
    \hline
  \end{tabular}
\end{table}
\begin{figure}[t]
	\centering
	\includegraphics[width=0.7\linewidth]{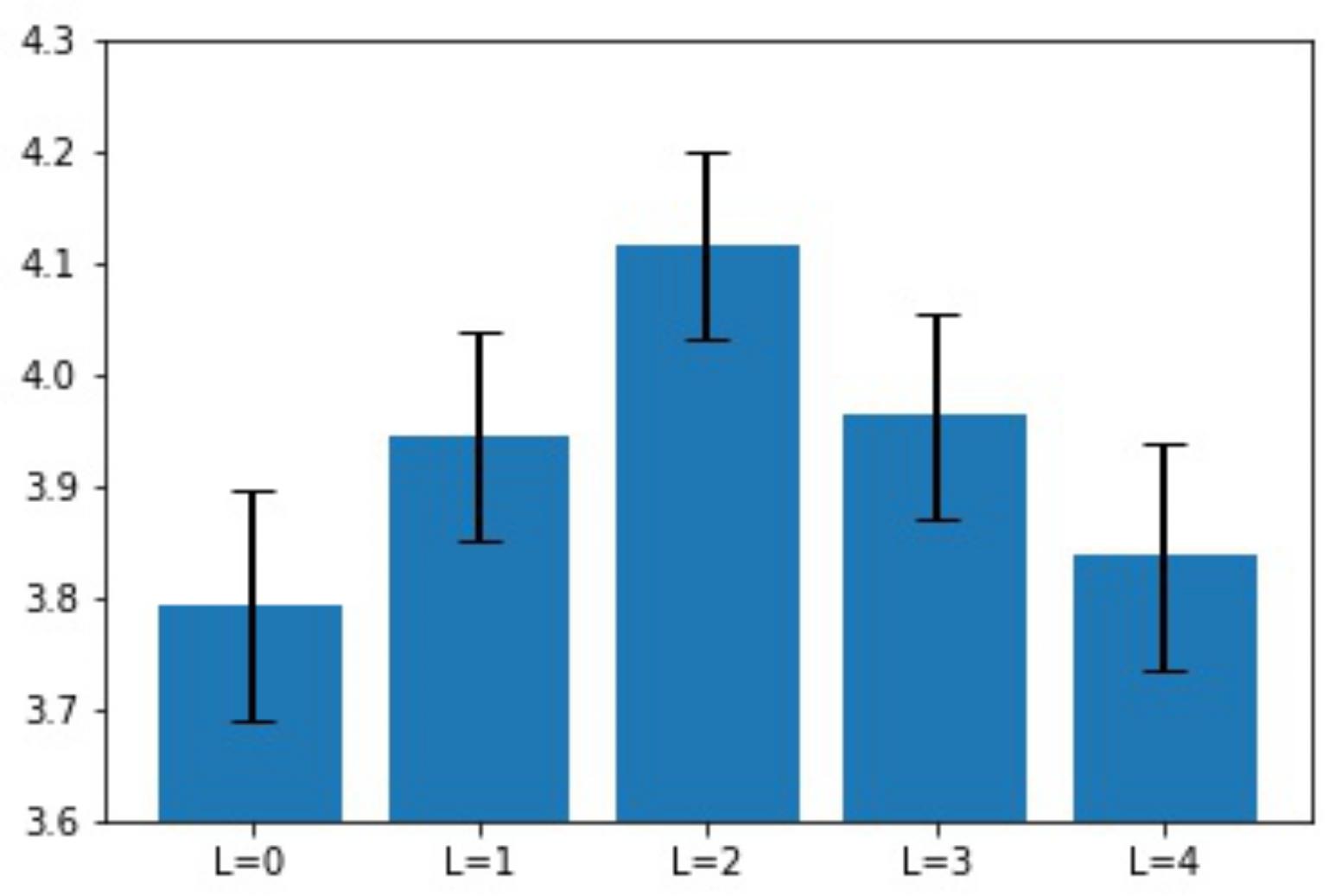}
	\caption{The MOS test scores on naturalness and expressiveness when using different ranges of context in predictor with 95\% confidence intervals.}
	\label{fig:mos}
\end{figure}
\begin{figure}[t]
	\centering
	\includegraphics[width=1\linewidth]{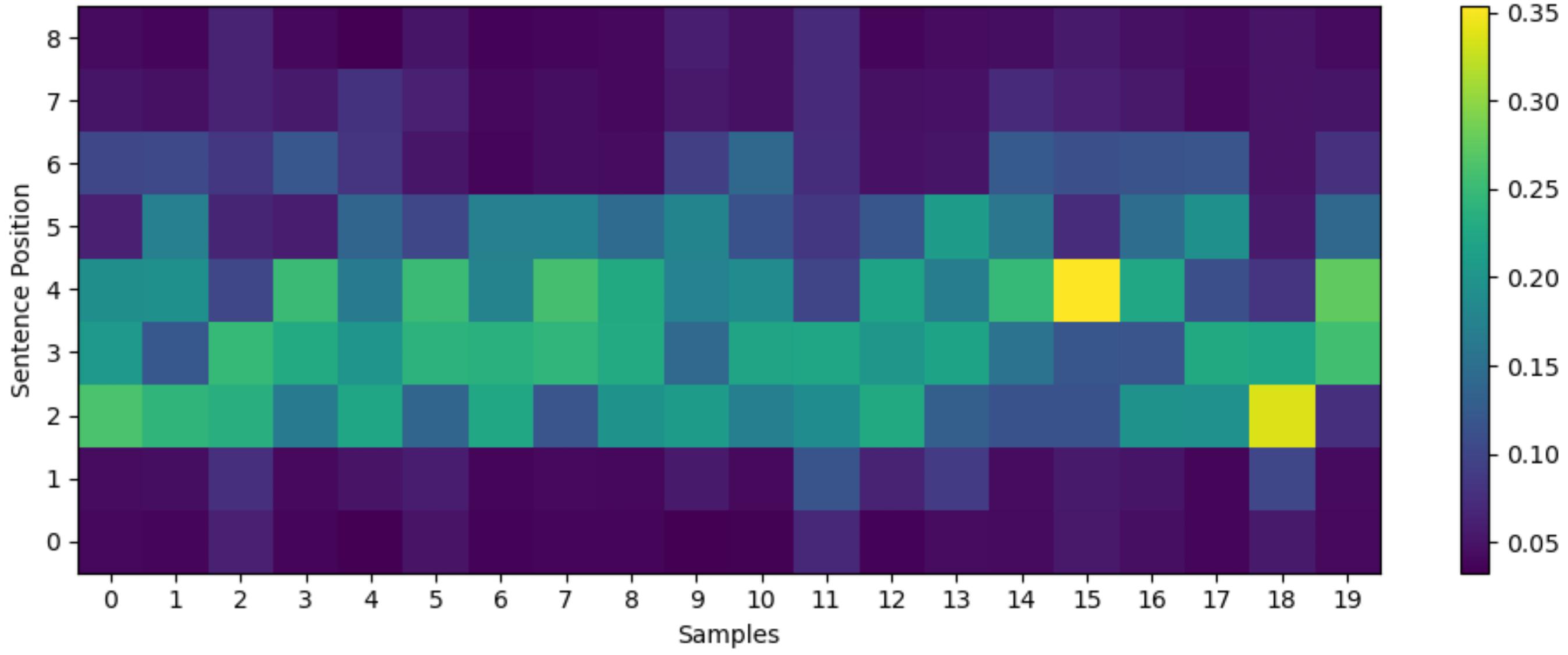}
	\caption{Attention weights of the inter-sentence level attention module in predictor when the number of sentences considered in the context is 9. The sentences in the context are numbered sequentially from 0 to 8 as vertical axis, where 0 means the first sentence in the context, 8 means the last sentence in the context, and 4 means the current sentence.}
	\label{fig:alignment}
\end{figure}
\begin{table*}[t]
  \caption{Objective evaluation results of ablation studies for the multi-scale style predictor and the style representations at different levels. Here, ``-residuals connections" stand for removing the residuals connections in the predictor, ``-global-level style" stand for ignoring the global-level style modeling, and ``-sentence-level style" stand for further ignoring the sentence-level style modeling on top of ignoring global-level style.}
  \label{tab:objpredictor}
  \centering
  \begin{tabular}{lccccc} 
    \hline
     \textbf{Models} &\textbf{MCD} & \textbf{F0 RMSE} & \textbf{Energy RMSE} & \textbf{Duration MSE} & \textbf{Style loss} \\
    \hline
    MSStyleTTS & $4.979$ & $62.544$ & $4.926$ & $0.201$ & $1.459$ ~~  \\
    \quad -residuals connections & $5.012$ & $62.350$ & $4.968$ & $0.203$ & $1.491$ ~~  \\
    \quad -global-level style & $5.037$ & $63.148$ & $4.975$ & $0.205$ & -~~  \\
    \qquad -sentence-level style & $5.181$ & $64.833$ & $5.201$ & $0.214$ & -~~  \\
    \hline
  \end{tabular}
\end{table*}
\begin{table}[h]
  \caption{Result of the ABX preference test between MSStyleTTS and the model without using residual connections in the predictor (-residual connections). N/P denotes ``no preference" and $p$ means the $p$-value of $t$-test between two models.}
  \label{tab:abxpredictor}
  \centering
  \begin{tabular}{cccc} 
    \hline
    \textbf{MSStyleTTS} & \textbf{-residual connections} & \textbf{N/P} &\textbf{$p$} \\
    \hline
    $52.6$ & $26.7$ & $20.7$ & \textless 0.001 ~~ \\
    \hline
  \end{tabular}
\end{table}
As introduced in Section \ref{sec:modelpredictor}, we predict multi-scale style from the context information that is composed of the current sentence, $L$ past sentences and $L$ future sentences.
By increasing or decreasing the number of sentences considered in the predictor, this experiment explores how the range of context influences the performance of the style modeling.
The objective evaluation results are shown in Table \ref{tab:objcontext}.
The model achieves the best performance across all evaluation metrics when $L = 2$ (i.e., the number of sentences considered in the context is 5).

Fig.\ref{fig:mos} further presents the MOS test results when different ranges of the context are used as input of the predictor.
First, we can see that all models perform well in terms of expressiveness, except the model considering only the text of the current sentence ($L=0$) and the model considering too many sentences in the context ($L=4$).
Second, the highest MOS result is obtained when the range of the context is 5 sentences ($L=2$) in comparison test, which is consistent with the objective evaluation results shown in Table \ref{tab:objcontext}.
As the number of sentences considered in the context decreases, the MOS of the synthesized speech decreases significantly, especially it shows the lowest MOS of $3.793$ when $L=0$. 
That is, the context information in the adjacent sentences is crucial for expressive speech synthesis, and increasing the range of context information can improve speech prosody.
However, we find that the expressiveness of the synthesized speech decreases with increasing contextual range when the number of sentences considered in the context is greater than 5.
It illustrates that considering the extra-long range of context information does not improve the perceived quality of the synthesized speech.
We suppose the reason is that the context with a long distance to the current sentence is not relevant for determining the style of the current.

To verify this hypothesis and visualize the contributions to style modeling of difference sentences that are of different distances to the current utterance, we randomly select 20 examples in the test set and plot the attention weights of the inter-sentence level attention module in the multi-scale style predictor when 9 sentences are considered in the context.
The result is shown in Fig.\ref{fig:alignment}.
As expected, although the input of the multi-scale style predictor contains 4 sentences in both the past and future context, the attention weights of the inter-sentence network tends to focus mainly on the past 2 sentences, the current sentence and the future 2 sentences (i.e., the sentence positions from 2 to 6).
That is, when sentences more than 5 are considered in the context, the extra context information does not sufficiently contribute to the style modeling, but increases the complexity of the modeling and affects the expressiveness of the synthesized speech.
This supports our hypothesis and explains why we only consider the past 2 sentences, the current sentence and the future 2 sentences in our implementation.

\subsubsection{The Effect of Multi-Scale Style Predictor}
The multi-scale style predictor consists of a hierarchical context encoder and a hierarchical style predictor.
Hierarchical context encoder is indispensable in the proposed multi-scale framework to extract the context information at different levels, and its effectiveness has been demonstrated in our previous work \cite{proposed}.
Therefore, we only evaluate the performance of the proposed hierarchical style predictor.
An ablation model has been built by removing the residual connections in this module.
The first two rows in Table \ref{tab:objpredictor} show the objective evaluation results of MSStyleTTS and the ablation model.
As can be seen, removing the residual connections causes worse MCD, RMSE of energy and MSE of duration.
This indicates the advantages of the residual connections in the hierarchical style predictor, which explicitly provides information about coarser-grained style when predicting the finer-grained style.
In addition, to measure the accuracy of style prediction, we calculate the MSE of style embedding between the predicted and the ground truth.
The results are shown in the last column in Table \ref{tab:objpredictor}.
With the residual connections in the predictor, MSStyleTTS has improved in style prediction, compared with the ablation model.
The ABX preference test is conducted between MSStyleTTS and the model without using residual connections in the hierarchical style predictor.
The results and the corresponding $p$-value are shown in Table \ref{tab:abxpredictor}.
The preference rate of our MSStyleTTS exceeds the ablation model by $25.9\%$, which demonstrates the effectiveness of our proposed hierarchical style predictor.

\subsubsection{Comparisons Between Global-Level, Sentence-Level and Subword-Level Style Representation} \label{sec:style}
\begin{table}[t]
  \caption{Results of the CMOS tests among the proposed model, the model without global-level style modeling and the model without both global-level and sentence-level style modeling.}
  \label{tab:cmos}
  \centering
  \begin{tabular}{l|c} 
    \hline
    \textbf{Models} &\textbf{CMOS} \\
    \hline
    MSStyleTTS & $0$ ~~~ \\
    \quad -global-level style & $-0.428$ ~~~ \\
    \qquad -sentence-level style & $-0.640$ ~~~ \\
    \hline
  \end{tabular}
\end{table}
In previous experiments, we have adopted two baseline models that consider mono-scale style modeling, including HCE for global-level style modeling and WSV* for word-level style modeling.
Comparison with these two models illustrates the advantage of multi-scale style modeling.
Although these two baseline models use the same context information and acoustic model as our proposed MSStyleTTS, they differ in extracting style representation and predicting style representation.

To further demonstrate the effectiveness of the global-level, sentence-level, and subword-level style representation, we first build an ablation model by removing the global-level style extraction module and prediction module.
The objective evaluation result is shown in the third row of Table \ref{tab:objpredictor}.
Our MSStyleTTS significantly outperforms the model without global-level style modeling, which indicates the effectiveness of modeling the global-level style variations for expressive speech synthesis.
Second, a new ablation model is built by further removing the sentence-level modules, that is, only modeling the subword-level style from context.
The last row in Table \ref{tab:objpredictor} shows the objective evaluation result of the new ablation model.
Comparing the results of these two ablation models, we can see that neglecting the sentence-level style modeling leads to a further decrease in all the metrics.

Table \ref{tab:cmos} shows the results of CMOS tests among our MSStyleTTS and the two ablation models mentioned above.
It can be seen that neglecting the global-level style affect the perceived quality of synthesized speech, while further neglecting the sentence-level style degrades the perceived quality of synthesized speech even more significantly.
These subjective evaluation results are consistent with the objective ones shown in Table \ref{tab:objpredictor} and again verify that in addition to fine-grained style modeling (e.g., word-level), coarse-grained style modeling (e.g., sentence-level and global-level) contributes equally to the perceived quality of the synthesized speech.  

\begin{figure}[t]
	\centering
	\includegraphics[width=1.0\linewidth]{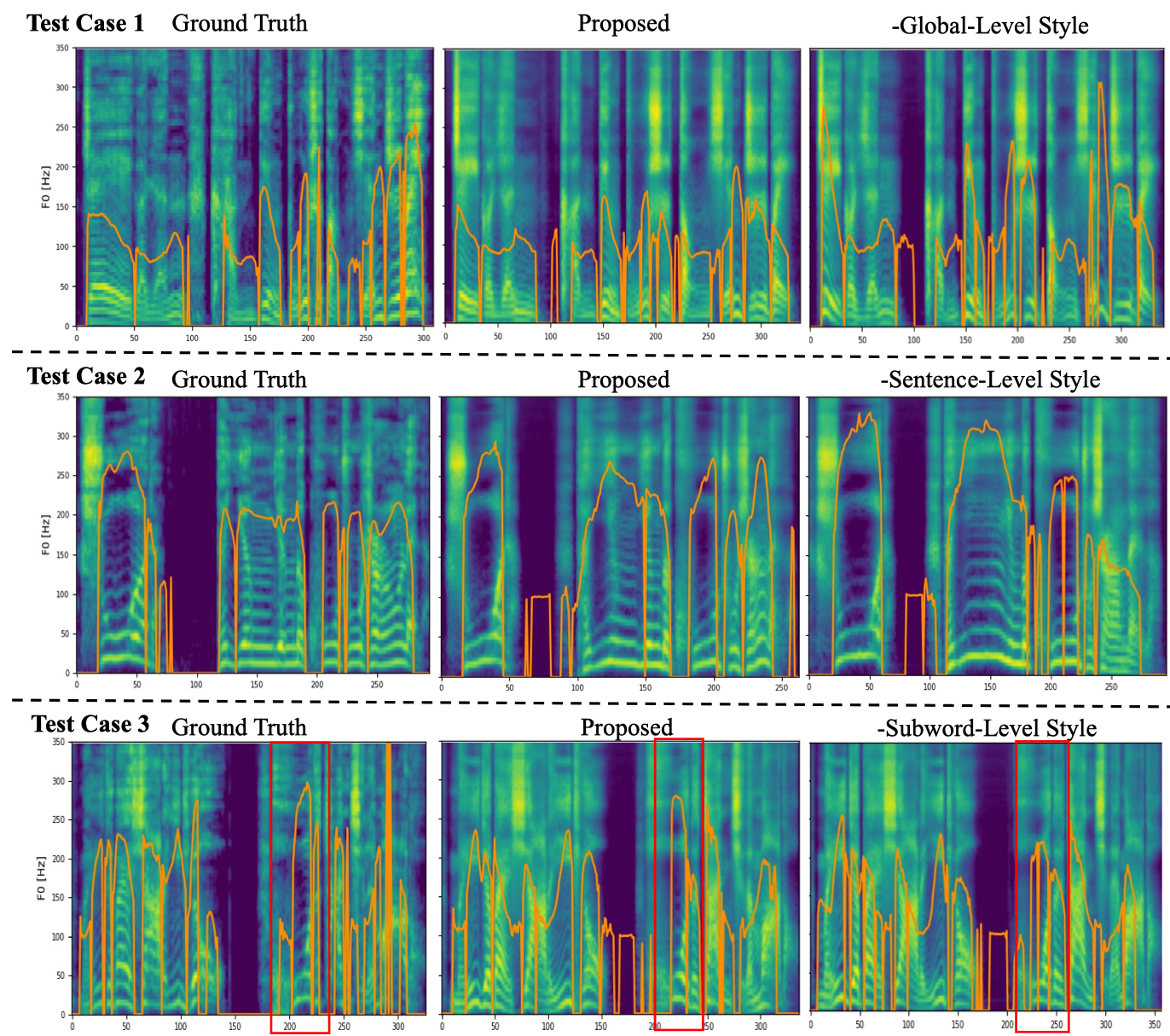}
	\caption{Mel-spectrograms and pitch contours of ground truth speeches, speeches synthesized with all three levels of style representations and speeches synthesized without each of the three levels of style representation for three test cases.}
	\label{fig:ablationcasestudy}
\end{figure}
\begin{table}[t]\scriptsize
  \caption{Objective evaluation on emotion classifiers of using style representation at different levels. ``Global-level" and ``Sentence-level" stand for using global-level or sentence-level style representation to train emotion classifier, respectively, ``Both" stand for using both the global-level and sentence-level style representation.}
  \label{tab:emotion}
  \centering
  \begin{tabular}{cccc}
    \hline
    \textbf{Granularity} & \textbf{Overall accuracy (\%)} & \textbf{F1 Weight (\%)} &\textbf{Kappa value} \\
    \hline
    Global-level & $82.77$ & $76.12$ & $0.0773$ ~~ \\
    Sentence-level & $81.55$ & $75.16$ & $0.0606$ ~~ \\
    Both & $83.54$ & $77.90$ & $0.1644$ ~~ \\
    \hline
  \end{tabular}
\end{table} 
To analyze what each level of style representation learns, we conduct three case studies to synthesize speeches without global-level, sentence-level, and subword-level style representations, respectively.
The style representation extracted from the ground truth speech is used in this evaluation.
We plot the mel-spectrograms and pitch contours of these speeches in Fig.\ref{fig:ablationcasestudy}, the ground truth speeches and speeches synthesized with all three levels of style representation are provided for comparison. 
It can be seen that removing the global-level or sentence-level style representation results in a higher overall pitch value,
which is inconsistent with the ground truth speech.
It indicates the ability of global-level and sentence-level style representation
to control the overall prosody of the sentence.
Moreover, the word enclosed by the red box is emphasized through a higher pitch value in the ground truth speech of test case 3.
However, the pitch contour of the word synthesized without subword-level style representation is much flatter than that of the ground truth and of the speech synthesized with all three levels of style representations.
It demonstrates that subword-level style representation can learn the prosody variations such as stress.

We have performed an emotion classification task on different levels of style representation to further analyze the effect of global-level and sentence-level style.
Table \ref{tab:emotion} shows the classification results.
As can be seen, the global-level style representation contains richer emotion-related information than the sentence-level style representation.
When utilizing both global-level and sentence-level style representation, the classifier has achieved the best results in terms of all evaluation metrics.
We conjecture that the possible reason may be the sentence-level style representation acts as supplement of the global-level one, which is more relevant to emotion strength and sentence-level prosody pattern.
\subsection{Discussion}
With the multi-scale style extractor and the multi-scale style predictor, our proposed method can synthesize speech with rich expressiveness by modeling multi-scale style from a wider range of context information. 

In the current method, the different levels of style representations are obtained from the ground truth speech in an unsupervised way.
Although the experiments in Section \ref{sec:style} illustrates that coarse-grained style representations contain emotion-related information, they do not perform well on the emotion classification task.
For future work, training the style extractor in a self-supervised manner is an alternative way to establish the connection between emotion and coarse-grained style representation without explicit emotion labels.
As for style prediction, it is worth exploring how to learn more efficient style representations with large amounts of speech data in an unsupervised manner.

It is worth noting that comparing the first and third rows in Table \ref{tab:objextractor}, there is still a large gap between the speeches synthesized based on the extracted style representations and the predicted style representations in terms of all evaluation metrics. 
This indicates that it is still challenging to model complex style variations in the real-world corpus.
In future work, we will explore a new model structure to further model speaking style by considering more contextual information.
\section{Conclusion} \label{sec:conclution}
This work proposes MSStyleTTS, an unsupervised multi-scale context-aware style modeling method for expressive speech synthesis.
MSStyleTTS allows modeling the speech style at different levels from hierarchical context information.
Experimental results show that our proposed method can synthesize speech with richer expressiveness and higher naturalness than baseline FastSpeech 2, WSV* and HCE models on both in-domain and out-of-domain datasets.
Extensive ablation studies demonstrate the effectiveness of the several techniques used in our proposed MSStyleTTS, especially for the context information usage and multi-scale style representation.

\bibliography{refs}
\bibliographystyle{ieeetr}

\begin{IEEEbiographynophoto}{Shun Lei} (Student Member, IEEE) received the B.E. degree from the School of Computer Science, Sichuan University, Chengdu, China, in 2021. He is currently working toward the M.E. degree in computer technology at Tsinghua University, China. His research interests include expressive speech synthesis, audiobook speech synthesis, 3D dance generation, and singing voice synthesis.
\end{IEEEbiographynophoto}
\begin{IEEEbiographynophoto}{Yixuan Zhou} (Student Member, IEEE) received the B.E. degree from the School of Information Science and Engineering, Southeast University, Nanjing, China, in 2020. He is currently working toward the Ph.D. degree in computer science and technology at Tsinghua University, China. His research interests include speech signal processing, text-to-speech synthesis, and few-shot/zero-shot learning.
\end{IEEEbiographynophoto}
\begin{IEEEbiographynophoto}{Liyang Chen} (Student Member, IEEE) received the B.E. degree in automation and the M.S. degree in control science from the School of Automation and Electrical Engineering, University of Science and Technology Beijing, Beijing, China, in 2018 and 2021 respectively. He is currently working toward the Ph.D. degree in computer science and technology at Tsinghua University, China. His research interests include speech signal processing, text-to-speech synthesis, and talking face generation.
\end{IEEEbiographynophoto}
\begin{IEEEbiographynophoto}{Zhiyong Wu} (Member, IEEE) received the B.S. and Ph.D. degrees in computer science and technology from Tsinghua University, Beijing, China, in 1999 and 2005, respectively. From 2005 to 2007, he was a Postdoctoral Fellow with the Department of Systems Engineering and Engineering Management, The Chinese University of Hong Kong (CUHK), Hong Kong. He then joined the Graduate School at Shenzhen (now Shenzhen International Graduate School), Tsinghua University, Shenzhen, China, and is currently an Associate Professor. He is also a Coordinator with the Tsinghua-CUHK Joint Research Center for Media Sciences, Technologies and Systems. His research interests include intelligent speech interaction, more specially, speech processing, audiovisual bimodal modeling, text-to-audio-visual-speech synthesis, and natural language understanding and generation. He is a Member of International Speech Communication Association (ISCA) and China Computer Federation (CCF).
\end{IEEEbiographynophoto}
\begin{IEEEbiographynophoto}{Xixin Wu} (Member, IEEE) received his B.S., M.S. and Ph.D. degrees respectively from Beihang University, Tsinghua University and The Chinese University of Hong Kong (CUHK). He is currently an Assistant Professor with the Department of Systems Engineering and Engineering Management, CUHK. Before this, he worked as a Research Associate with the Machine Intelligence Laboratory, Engineering Department of Cambridge University, and a Research Assistant Professor at the Stanley Ho Big Data Decision Analytics Research Centre, CUHK. His research interests include speech synthesis and recognition, speaker verification, and neural network uncertainty. He is a Member of International Speech Communication Association (ISCA) and China Computer Federation (CCF).
\end{IEEEbiographynophoto}
\begin{IEEEbiographynophoto}{Shiyin Kang} (Member, IEEE) received the B.S. and M.S. degrees from Tsinghua University, and the Ph.D. degree from The Chinese University of Hong Kong. He is currently a research director at Kunlun Group, where he is leading the audio generation team of the Skywork AI Music department. He has led several pioneering R\&D projects in Tencent AI Lab, Huya, and XVerse Inc., including Multi-Modal Text-to-Speech Synthesis Platform, AI singer, AI dancer, live show captioning, etc. His research interests include speech synthesis, singing voice synthesis, voice conversion, speech recognition, music audio generation, facial-expression, gesture and dancing generation for digital human, and applications in machine learning. He is a Member of International Speech Communication Association (ISCA) and China Computer Federation (CCF).
\end{IEEEbiographynophoto}
\begin{IEEEbiographynophoto}{Helen Meng} (Fellow, IEEE) received the B.S., M.S., and Ph.D. degrees in electrical engineering from the Massachusetts Institute of Technology, Cambridge, MA, USA. In 1998, she joined The Chinese University of Hong Kong, where she is currently a Patrick Huen Wing Ming Professor with the Department of Systems Engineering and Engineering Management. She was the former Department Chairman and the Associate Dean of Research with the Faculty of Engineering. Her research interests include human-computer interaction via multimodal and multilingual spoken language systems, spoken dialog systems, computer-aided pronunciation training. She was the Editor-in-Chief of the IEEE Transactions on Audio, Speech and Language Processing between 2009 and 2011. She was the recipient of the IEEE Signal Processing Society Leo L. Beranek Meritorious Service Award in 2019. She was also an Elected Board Member of the International Speech Communication Association (ISCA) and an ISCA International Advisory Board Member. She is a Fellow of the ISCA, HKCS, and HKIE.
\end{IEEEbiographynophoto}
\vfill
\end{document}